\documentclass[aps,prl,twocolumn,groupedaddress,showpacs]{revtex4-1}
\pdfoutput=1

\usepackage{amsmath}
\usepackage{amssymb}
\usepackage{graphicx}
\usepackage{bm}
\usepackage{times}

\usepackage{hyperref}
\hypersetup{%
  breaklinks = {true},
  citecolor = {blue},
  colorlinks = {true},
  linkcolor = {red},
  pdfauthor = {\textcopyright\ Johan Nilsson},
  pdfcreator = {\LaTeX\ and \flqq hyperref\frqq},
  pdffitwindow = {true},
  pdfmenubar = {true},
  pdfpagelayout = {SinglePage},
  pdfstartview = {Fit},
  pdftoolbar = {true},
  plainpages = {false},
}

\newcommand{\la}{\langle}
\newcommand{\ra}{\rangle}

\AtBeginDocument{\providecommand{\hslash}{\hslash}}

\usepackage{color}

\begin{document}

\pacs{
03.65.Ge,	
03.65.Sq,	
73.22.Pr	
}

\title{Trapping massless Dirac particles in a rotating saddle}

\author{Johan Nilsson}
\affiliation{Department of Physics, University of Gothenburg, 412 96 Gothenburg,  Sweden}
\email{johan.nilsson@physics.gu.se}

\date{July, 2013}

\begin{abstract}

We study particle motion in rotating saddle-shaped potentials. It is known that such rotating potentials can generate bounded motion for particles with a parabolic dispersion law through the combination of potential, centrifugal and Coriolis forces in the rotating frame. When applied to massless Dirac particles, for example electrons in graphene, such a potential is shown to lead to eigenstates that are spatially localized near the center of the saddle at certain energies. Although other states also exist at these energies, they have non-overlapping support in the oscillator basis, which tend to give the localized states a substantial life-time also when imperfections are present.

\end{abstract}

\maketitle

It is well-known that it is impossible to trap a charged particle with static external electric fields in three dimensions. Allowing for magnetic fields or time-dependent potentials changes this picture. Indeed, in atomic physics ions are routinely trapped and studied in Penning \cite{GabrielseRMP,DehmeltRMP1990} and Paul \cite{PaulRMP1990} traps. In a Penning trap the combination of an external perpendicular magnetic field and a repulsive potential generate bounded motion in two dimensions (2D); when a particle tries to escape outwards it is deflected by the Lorentz force and pushed into bounded orbits. In the Paul trap a periodically driven quadrupole field is used to confine the ions in 2D. The physics of the Paul trap is often motivated by the closely related mechanical problem of a point-like particle that is moving in two dimensions in a rotating saddle-shaped potential \cite{PaulRMP1990,BirulaPRL1994,Canadian2002}. This problem (both the classical and the quantum version) can be solved exactly with elementary methods, as will be outlined below. In contrast the analysis of the Paul trap rests on the solutions of the Mathieu equation, which is considerably more involved. The relation between the rotating saddle potential and the Paul trap is akin to the relation between the Jaynes-Cummings model \cite{JaynesCummings1963} and the Rabi model \cite{Rabi1937}, in that the exact solution of the first is elementary while the analysis of the second one is considerably more involved \cite{BraakPRL2011}. As we shall see, the analysis of our system is greatly simplified by the fact that it is effectively time-independent in the rotating frame.

A related and important problem in the field of graphene is the great difficulty of creating bound states by electrostatic means. The reason for this difficulty is that both electrons and holes have to be confined at the same time, but this is not possible in the absence of a (band) gap. Apart from the special case exactly at zero energy, where algebraically decaying states may be found  \cite{VitorPRL2006,JensBoundstates2009}, essentially all states are extended scattering states, for a possible exception see \cite{JensBoundstates2009}. Sticking to static electric fields, atomic-scale potentials are generally believed to be necessary for confinement. Classification of possible mass terms in graphene can be found in \cite{GrapheneMasses2009,Herbut2012a}, and examples that generate bulk gaps include static staggered sublattice potentials \cite{Semenoff1984}, antiferromagnetism \cite{Sorella1992}, and spin-orbit coupling \cite{KaneMeleQSH1}. It is however presently not possible to have sufficient control over these gaps. It is of course also possible to cut out a small flake and terminate the crystal. This creates confinement, but at the expense that the intriguing Dirac fermion physics might be lost due to lattice effects. It would be better if smooth potentials could be used to create confinement.

In this letter we present a scheme to trap electrons in graphene (and other particles described by a quasi-relativistic massless 2D Dirac equation) by subjecting them to a rotating saddle-shaped potential. This can be achieved by having two superimposed time-periodic quadrupole potentials with the proper rotation and phase shift. Our proposal is a natural extension of the recently popularized idea to modify material properties through spatially homogeneous periodic driving \cite{Syzranov2008,Kitagawa2011,DrivenMajorana2011,Cayssol2013,DrivenKekule}, to also allow for spatially dependent periodic driving. We present theoretical arguments for trapped electron states by studying the Dirac equation in a rotating potential using semi-classical analysis, perturbative arguments, and a numerical investigation of the full quantum problem. To set the stage we first briefly review the behavior of the related quadratic problem, since this can be solved exactly and the result might be unfamiliar to non-experts.

{\it Classical particle}. Let us begin by studying a classical particle with mass $m$ that moves in two dimensions in a saddle-shaped potential
\begin{equation}
V(\mathbf{x}) =  \frac{m}{2} (\omega^2_1 x_1^2 -\omega^2_2 x_2^2 ),
\label{eq:Vrotatingsaddle}
\end{equation}
that is rotating with angular frequency $\Omega$ with respect to the lab frame. $\omega_1$ and $\omega_2$ are frequencies that parametrize the shape of the potential. The classical equations of motion in the rotating frame are 
\begin{equation}
\begin{split}
\ddot{x}_1  & = \Omega^2 x_1 + 2\Omega \dot{x}_2 -\omega_1^2 x_1 , \\
\ddot{x}_2  & = \Omega^2 x_2 -2\Omega \dot{x}_1 +\omega_2^2 x_2  .
\end{split}
\end{equation}
On the right-hand side the first term is due to the centrifugal force and the second to the Coriolis force \cite{GoldsteinBook1980}. Solutions are easily found through the ansatz $x_1 = c_1 e^{-i \lambda t}$ and $x_2 = c_2 e^{-i \lambda t}$ with $c_1$ and $c_2$ constants. $\lambda$ is determined from the solutions of the characteristic equation, which can be expressed as $\lambda^2 = (\Omega \pm \Omega_1)^2 - \Omega_2^2$, where $\Omega_2 = (\omega_1^2 + \omega_2^2)/(4 \Omega)$ and $\Omega_1 = \sqrt{(\omega_1^2 - \omega_2^2)/2 +\Omega_2^2}$. For stability all $\lambda$ must be purely real, this gives two cases
\begin{equation}
\begin{split}
\label{eq:quadraticstability}
\omega_2^2 & \leq \omega_1^2 \leq \Omega^2   \quad \text{or} \\
\omega_1^2 & \leq \omega_2^2 \leq 3 \omega_1^2 
\quad  \text{and} \quad 
\omega_1^2 \leq \Omega^2 \leq \frac{(\omega_1^2+\omega_2^2)^2}{8 (\omega_2^2 - \omega_1^2)} .
\end{split}
\end{equation}
This means that when the saddle is more attractive than repulsive the driving frequency has to be large enough to achieve stability, while if the saddle is more repulsive than attractive and not too repulsive there is a finite frequency interval where stable motion is possible.


{\it Quantum particle with parabolic dispersion}. Let us now consider a quantum particle with a quadratic dispersion relation that moves in 2D  in the same rotating potential as in the classical case \eqref{eq:Vrotatingsaddle}. In the rotating frame the stationary states are given by the solutions to the eigenvalue problem
\begin{equation}
H_p  \Phi = 
\frac{\mathbf{p}^2}{2m}   \Phi +V(\mathbf{x})  \Phi
- \Omega L_3  \Phi
\\ =  E \Phi .
\label{eq:Hamiltonian_rotatingframe1}
\end{equation}
Here $L_3 = x_1 p_2 - x_2 p_1$ is the usual angular momentum operator. This Hamiltonian can be diagonalized by a Bogoliubov transformation when the motion is stable \cite{Hasegawa2010,supplementalBogoliubov}, the result is
\begin{equation}
H_p = \hslash \lambda_+ (a^{\dagger}_+ a^{\,}_+ -1/2) 
- \hslash \lambda_- (a^{\dagger}_- a^{\,}_- - 1/2),
\end{equation}
with $\lambda_\pm  = \sqrt{ (|\Omega| \pm \Omega_1)^2 - \Omega_2^2}$. Here $a_\pm$ are two independent bosonic annihilation operators. Stable motion is possible only when $\lambda_\pm$ are both real, which gives the same conditions as for the classical case \eqref{eq:quadraticstability}. The complete set of square integrable eigenstates are the usual ladder of harmonic oscillator eigenstates built upon the vacuum annihilated by both $a_+$ and $a_-$. The fact that the energy eigenvalues are not bounded from below makes the system meta-stable in the presence of perturbations, the life-time can nevertheless be quite long \cite{BirulaPRL1994}.

{\it Semi-classical treatment of massless Dirac particles}. After this introduction we now consider the 2D massless Dirac Hamiltonian in an external potential
\begin{equation}
i \hslash \frac{d}{dt} \Psi_{lab} = 
v  \mathbf{p}_{lab} \cdot \bm{\sigma}  \Psi_{lab}
+ V(\mathbf{x}_{lab},t) \Psi_{lab} ,
\end{equation}
here $v > 0$ is  the velocity of the particles. In the case of graphene this is the Hamiltonian close to one of the K-points and the Pauli matrices $\bm{\sigma}$ act in sublattice space \cite{AntonioRMP}. Stationary states can be found by going to the rotating frame
$\Psi_{lab}(t,\mathbf{x}_{lab}) =  e^{-i E t / \hslash} e^{-i (\Omega t /2) \sigma_3} \Psi (\mathbf{x} )$,
which leads to the eigenvalue problem
\begin{equation}
H \Psi = 
v  \mathbf{p} \cdot \bm{\sigma}  \Psi
+ V(\mathbf{x}) \Psi - \Omega J_3 \Psi  = E \Psi.
\end{equation}
Here $J_3 = L_3 + \hslash \sigma_3 /2$ is the total angular momentum operator, which has eigenvalues that are half-integer multiples of $\hslash$: $\pm
\hslash/2, \pm 3 \hslash /2, \pm 5\hslash /2,$ etc. Since there is no mass in the bare graphene problem we define a dynamically generated mass via $m v^2 = \hslash |\Omega| / 2 $ and continue to use the parametrization of the potential in \eqref{eq:Vrotatingsaddle}. Let us now perform a semi-classical analysis of this Hamiltonian. Diagonalizing the matrix structure we get two semi-classical energy bands corresponding to electrons ($+$) and holes ($-$): $E_{\pm} = \pm E_{\mathbf{p}} - \Omega L_3 + V(\mathbf{x})$,
with $E_{\mathbf{p}} = \sqrt{v^2 \mathbf{p}^2 + (m v^2)^2}$. The semi-classical equations of motion for wave-packets in multi-band systems such as this one are modified by a Berry curvature term \cite{SundaramNiu1999,SemiclassicalBerryRMP2010}
\begin{equation}
\begin{split}
\dot{\mathbf{x}}  &=\nabla_{\mathbf{p}} E_{\pm} 
 - \dot{\mathbf{p}} \times \mathbf{\Omega}_{\pm}(\mathbf{p}) /\hslash , \\
\dot{\mathbf{p}}  &= -\nabla_{\mathbf{x}} E_\pm .
\end{split}
\end{equation}
In our case the Berry curvature is
$\mathbf{\Omega}_\pm = \pm \hslash^3 v^2 \Omega  \hat{n}_3 / (4 E_{\mathbf{p}}^3)$.
Eliminating $\dot{\mathbf{x}}$ from the equations of motion we arrive at
\begin{equation}
\begin{split}
\ddot{p}_1  & = \frac{\Omega (\omega_2^2- \omega_1^2)}{\omega_2^2} \dot{p}_2
- \frac{\Omega^2 \omega_1^2}{\omega_2^2} p_1
\mp \frac{\hbar |\Omega| \omega_1^2}{2 E_{\mathbf{p}}} p_1
\pm \frac{\hslash^3 \Omega | \Omega | \omega_1^2}{8 E^3_{\mathbf{p}}} \dot{p}_2, \\
\ddot{p}_2  & = \frac{\Omega (\omega_2^2- \omega_1^2)}{\omega_1^2} \dot{p}_1
- \frac{\Omega^2 \omega_2^2}{\omega_1^2} p_2
\pm \frac{\hbar |\Omega| \omega_2^2}{2 E_{\mathbf{p}}} p_2
\pm \frac{\hslash^3 \Omega | \Omega | \omega_2^2}{8 E^3_{\mathbf{p}}} \dot{p}_1 ,
\end{split}
\end{equation}
which determines the semi-classical wave-packet dynamics for electrons (upper signs) and holes (lower signs).
For $|\mathbf{p}| \ll m v$ we can linearize these equations and look for stable solutions exactly as for the classical particle above, i.e., we make the ansatz $p_1 = c_1 e^{-i\lambda t }$ and $p_2 = c_2 e^{-i\lambda t }$ and solve for $\lambda$. Note that the linearization implies that $E_{\mathbf{p}} \approx \hslash |\Omega|/2 = mv^2$. Demanding that all of the eigenvalues are real for wave-packets constructed in both of the bands we get the necessary condition $(\sqrt{2}-1/2) \omega_>^2 < \omega_<^2  \leq \omega_>^2$ where $\omega^2_{>}$ ($\omega^2_{<}$)  is the greater (smaller) of $\omega_1^2$ and $\omega_2^2$. As a result the two curvatures $\omega^2_1$ and $\omega_2^2$ are not allowed to be too different. In addition there is the constraint $\omega_>^2 \leq \Omega^2 \leq  \omega_>^2 f_c (|\omega_< /\omega_>|)$, with $f_c(x) =  x^2 / \{ (1-x^2)^{1/3}[(1+x)^{2/3}-(1-x)^{2/3}]^2\}$. This implies that if $\omega_1^2 \neq \omega_2^2$ the motion is only stable for a finite interval of driving frequencies $\Omega$. Specializing to the particle-hole symmetric case (see discussion below) where $\omega^2_1 = \omega^2_2 = \omega^2$ the solutions are $\lambda^2 = \Omega^2$ and $\lambda^2 = \Omega^2 - \omega^4 / \Omega^2$, and hence a sufficient condition for stability in this case is simply $\Omega^2 \geq \omega^2$. Although this analysis hints at what to expect from a full quantum mechanical solution it should be viewed with some skepticism. For the semi-classical analysis to be valid the time scale associated with the wave-packet dynamics should be much longer than the time-scale of the gap, i.e.,  $|\lambda| \ll |\Omega|$. This is not true for all of the semi-classical bounded solutions, although it is for some.


{\it Quantum treatment of particles with linear dispersion}. 
To study the quantum problem it is convenient to first reformulate it in dimensionless variables.  Denoting $\omega^2 = (\omega_1^2 + \omega_2^2)/2$ and $\Delta^2 = \omega_1^2 - \omega_2^2$  we introduce the characteristic length scale  $l =2v / \omega$ and write  $ x_j/l = \tilde{x}_j$. We also go to a different gauge $\Psi \rightarrow \exp(i  s_\Omega \tilde{x}_1 \tilde{x}_2) \Psi $ with $s_\Omega = \text{sign}(\Omega)$ which takes $\tilde{p}_1 \rightarrow \tilde{p}_1 + s_\Omega  \tilde{x}_2$  and $\tilde{p}_2 \rightarrow \tilde{p}_2 + s_\Omega \tilde{x}_1$ so that the Hamiltonian goes into
\begin{equation}
\frac{H}{\hslash \Omega} = - \tilde{J}_3
+ \frac{\omega }{2 \Omega}
( \tilde{\mathbf{p}} \cdot \bm{\sigma} + i s_\Omega \tilde{\mathbf{x}} \cdot \bm{\sigma}^* \sigma_3 )
+ \frac{s_\Omega \Delta^2}{2 \omega^2 } 
(\tilde{x}_1^2+\tilde{x}_2^2).
\end{equation}
We now reexpress $H$ in an oscillator basis where $\tilde{x}_j$ and $\tilde{\partial}_{j}$ are represented with creation and annihilation operators in the usual way $a_j^{\,}   = (\tilde{\partial}_{j} +  \tilde{x}_j)/\sqrt{2} $. In terms of the bosons $b_1^{\,} = (-i a_1^{\,} + a_2^{\,})/\sqrt{2}$ and $b_2^{\,} = (-i a_1^{\,} - a_2^{\,})/\sqrt{2}$ we get
\begin{multline}
 \frac{H}{\hslash \Omega} =
-\tilde{J}_3
+
\frac{\omega}{2 \Omega} 
\begin{pmatrix}
0 & b_1^\dagger + b_2^{\,} \\
\text{h.c.} & 0
\end{pmatrix}
+  \frac{s_\Omega \Delta^2 }{2 \omega^2 } \tilde{r}^2
\\
+ 
 \frac{s_\Omega  \omega}{2 \Omega} 
\begin{pmatrix}
0 & b_1^{\,}   -b_2^\dagger  \\
\text{h.c.} & 0
\end{pmatrix},
\label{eq:fullQHamiltonian}
\end{multline}
with
\begin{equation}
\begin{split}
\tilde{J}_3   &=  b_2^\dagger b_2^{\,} - b_1^\dagger b_1^{\,} + \sigma_3 /2, \\
\tilde{r}^2   &= 
\tilde{x}_1^2+\tilde{x}_2^2 = b_1^\dagger b_1^{\,} + b_2^\dagger b_2^{\,} + 1 - b_1^{\,} b_2^{\,} - b_2^\dagger b_1^\dagger  .
\end{split}
\end{equation}
When $\Delta^2 = 0$ the spectrum is particle-hole symmetric. This can be implemented by the transformation $b_1 \rightarrow i b_2$ and $b_2 \rightarrow - i b_1$ combined with the rotation $R =e^{-i \pi \sigma_3 /4} \sigma_2 $ on the Hamiltonian \eqref{eq:fullQHamiltonian}, this also takes $\tilde{J}_3 \rightarrow -\tilde{J}_3$.

We note that all of the terms on the first line of \eqref{eq:fullQHamiltonian} commutes with $\tilde{J}_3$ which has half-integer eigenvalues $j = \pm1/2,\pm 3/2, \ldots$, whereas the term on the second line couple states with values $j$ to $j \pm 2$. If the first term is the largest one a straightforward degenerate perturbation theory calculation to second order in the remaining terms, similar to the one in \cite{Kitagawa2011},  gives an effective Hamiltonian that commutes with $\tilde{J}_3$
\begin{equation}
 \frac{H_{\text{eff}}}{\hslash \Omega} =
-\tilde{J}_3
+
\frac{\omega}{2 \Omega}
 \tilde{\mathbf{p}} \cdot \bm{\sigma}
-  \Bigl( \frac{ \omega}{2 \Omega}  \Bigr)^2
\frac{\tilde{r}^2}{2} \sigma_3
+  \frac{s_\Omega \Delta^2}{2 \omega^2 } \tilde{r}^2 .
\label{eq:Heff}
\end{equation}
This is a 2D Dirac equation with a spatially varying mass term, which is an established way to confine Dirac particles \cite{BerryMondragon1987}. From $H_{\text{eff}}$ we also get a simple necessary condition for confinement by demanding that the strength of the mass term is larger than that of the parabolic potential. This implies that $\Omega^2 |\Delta^2| < \omega^4 / 4 $ and puts an upper bound on $\Omega^2$ that goes to infinity in the limit $\Delta^2 \rightarrow 0$. On the other hand we also need $\omega \ll |\Omega|$ for the perturbative treatment to be valid. This result is therefore consistent with the semi-classical stability analysis above with an upper and lower bound on the allowed driving frequency, but less numerically precise. Solving $H_{\text{eff}}$ numerically for $j=-1/2$ we find that the perturbative treatment is self-consistent for states with energies close to $\hslash \Omega/2$. 

\begin{figure}
\centering
\includegraphics[scale=.45]{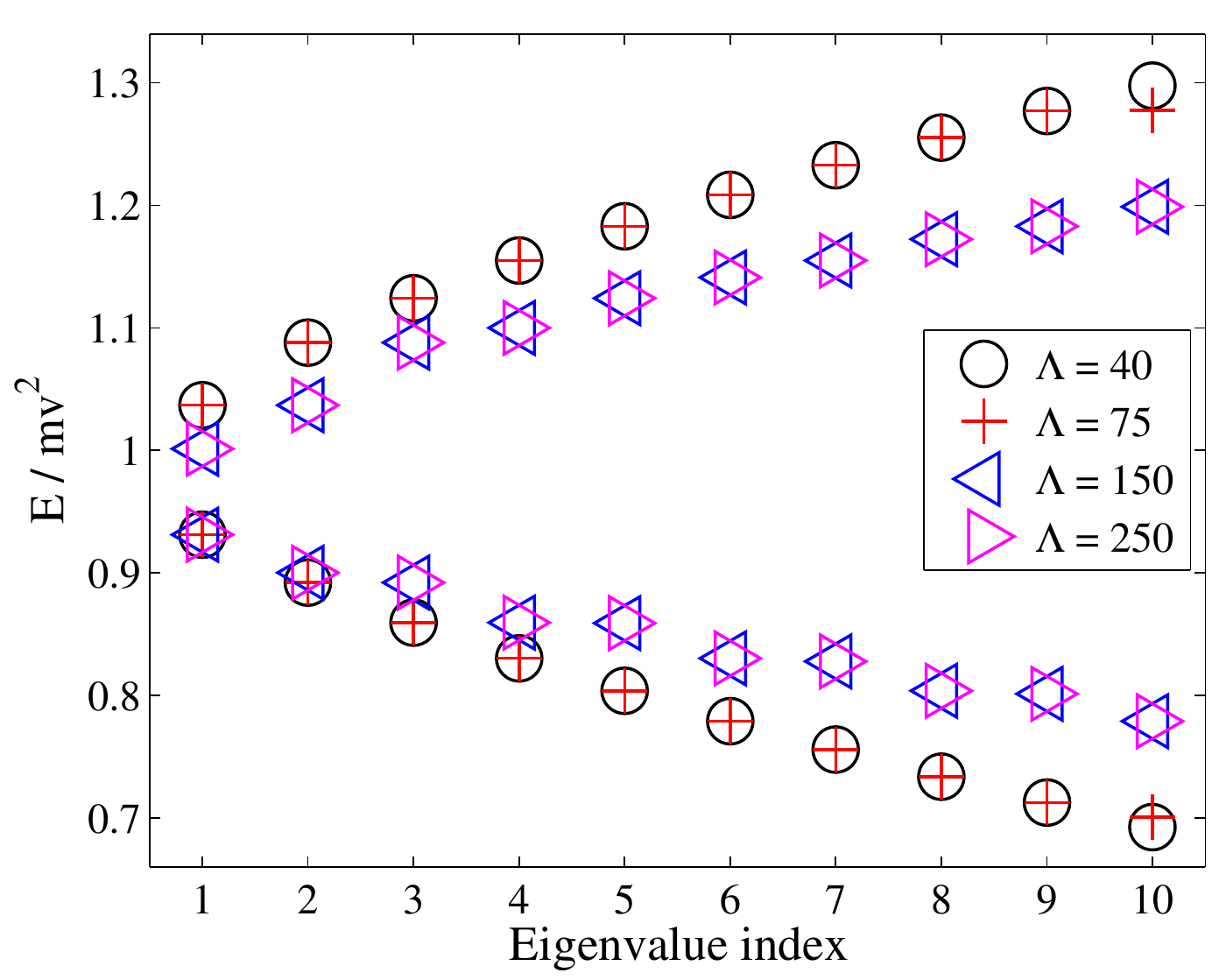}
\caption{Normalized energy eigenvalues of $H / mv^2$ closest to $1$ obtained by numerical diagonalization of the full particle-hole symmetric quantum problem for $\omega / \Omega = 0.1$ and different cut-offs $\Lambda$. The convergence with the number of $j$-states kept is rather quick for these parameters, 
here we use 21 states centered around $j=-1/2$.
Some eigenvalues, corresponding to $j = -1/2$, converge and remain stable as $\Lambda$ is increased. For sufficiently large $\Lambda$ ($\sim 100$ in this case) new states with energies near $E/mv^2 = 1$ appear and that are also stable upon further increase of the cut-off. These have a large overlap with states of angular momentum $-1/2 \pm 1$.
\label{fig:eigenvalues}}
\end{figure}

\begin{figure}
\begin{center}
\includegraphics[scale=.5]{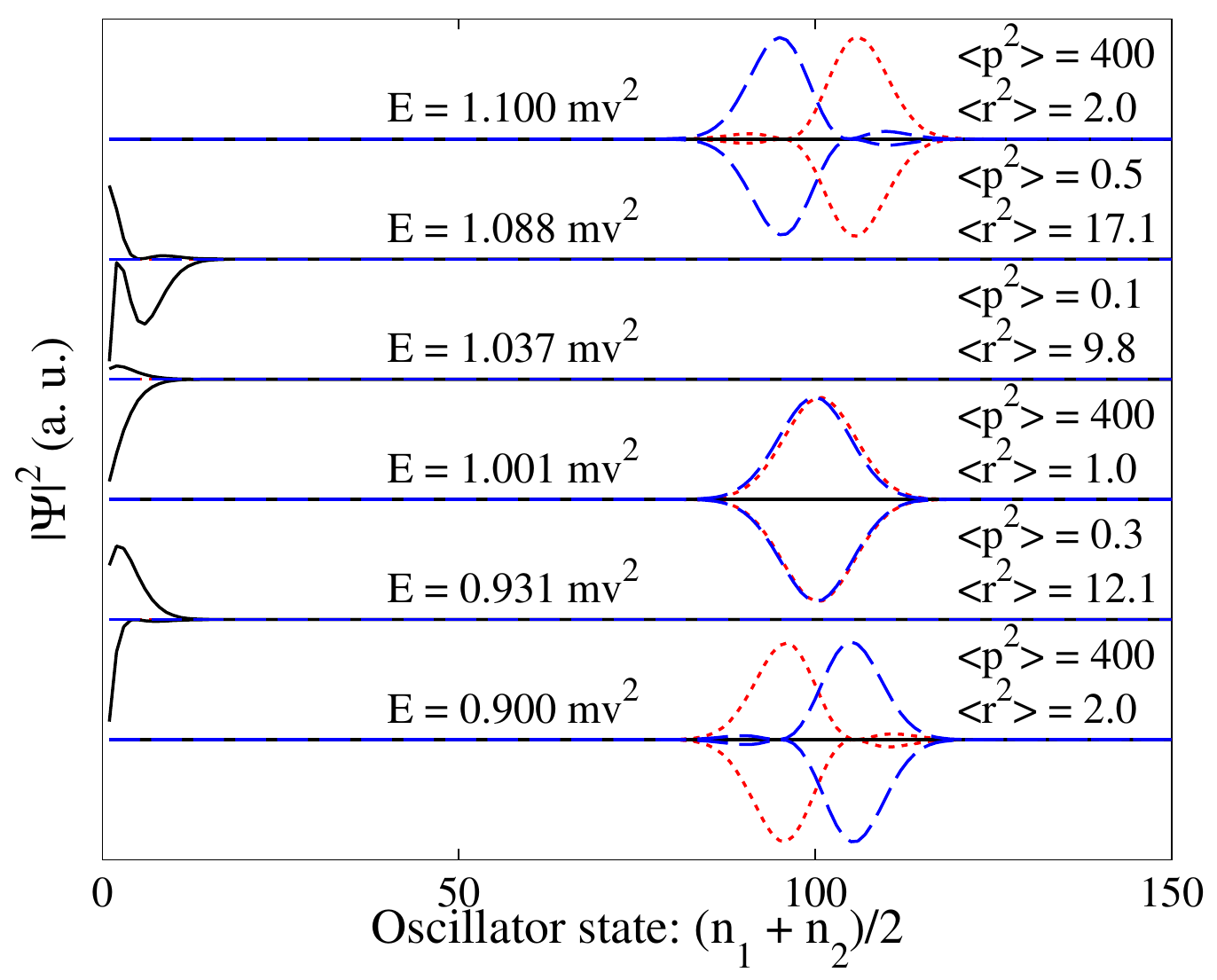}
\end{center}
\caption{Weights in different oscillator states for some of the eigenfunctions corresponding to Fig.~\ref{fig:eigenvalues}  ($\Lambda = 10^3$).
Weights for different levels are shifted and rescaled independently and inverted on one of the sublattices.  The curves denote: $j=-1/2$ (black solid line); $j=-3/2$ (red dotted line); $j=1/2$ (blue dashed line). The weight in the other angular momentum channels is very small. We also indicate the value of $\la \tilde{r}^2 \ra$ and $\la \tilde{p}^2 \ra$.
\label{fig:wavefunctions}}
\end{figure}

For simplicity we now consider the particle-hole symmetric case $\Delta^2 = 0$ and that $\Omega \gg \omega$. When $\omega = 0$ the eigenenergies are $-\hslash \Omega j$ with $j = \pm1/2,\pm 3/2, \ldots$, all of which are infinitely degenerate. Perturbatively the degeneracy is split when $\omega \neq 0$. It is straightforward to write down the matrix representing the Hamiltonian  \eqref{eq:fullQHamiltonian} in the standard occupation number basis on each sub-lattice: $\{ |n_1 , n_2 \ra \}_{n_1, n_2 \geq 0}$.
We truncate the matrix by keeping a finite range of angular momentum states $j_{min} \leq j \leq j_{max}$, and for each $j$-value we keep the $\Lambda$ states with the lowest possible values of $n_1 + n_2$. Diagonalizing this matrix numerically we find that some eigenstates and the associated energies converge near $E \approx \hslash \Omega/2$, see Figs.~\ref{fig:eigenvalues}-\ref{fig:wavefunctions}.
These are the localized states of main interest, and they have a large weight in the $j=-1/2$ sector. There are also other states near this energy, but they are constructed from large values of $n_1$ and $n_2$.  Increasing the cutoff $\Lambda$, the first such states that appear have large roughly equal weights in the $j = -1/2 \pm 1$ sectors. These states are therefore not described by the effective Hamiltonian \eqref{eq:Heff}. A simple estimate of the associated quantum numbers is obtained by setting $b_1 \sim \sqrt{n_1} \sim b_2 \sim \sqrt{n_2}$ and therefore $n_1 \sim n_2 \sim (\Omega/\omega)^2$. This estimate is consistent with the numerically obtained wave-functions shown in Fig.~\ref{fig:wavefunctions}. Clearly more states appear when larger cut-offs $\Lambda$ are considered corresponding to even bigger $|j|$. An important observation is that since the different sets of wavefunctions  have essentially non-overlapping support, smooth disorder potentials will only weakly couple the different sets, which implies that the life-times of the localized states are large. To characterize the eigenstates further we also calculate expectation values of $\la \tilde{r}^2 \ra$ and $\la \tilde{p}^2 \ra$, see Fig.~\ref{fig:wavefunctions}. Since $\la \tilde{r}^2 \ra$ is finite the states are indeed localized. The character of the eigenstates with weights in the  $j = -1/2 \pm 1$ sectors is also clarified by these quantities: they are localized close to the origin but with large momentum to compensate for the energy difference due to $\tilde{J}_3$. The only states with both $\la \tilde{r}^2 \ra, \la \tilde{p}^2 \ra \lesssim |\Omega|/\omega$ are those corresponding to \eqref{eq:Heff}.

\begin{figure}[h!]
\begin{center}
\includegraphics[scale=.9]{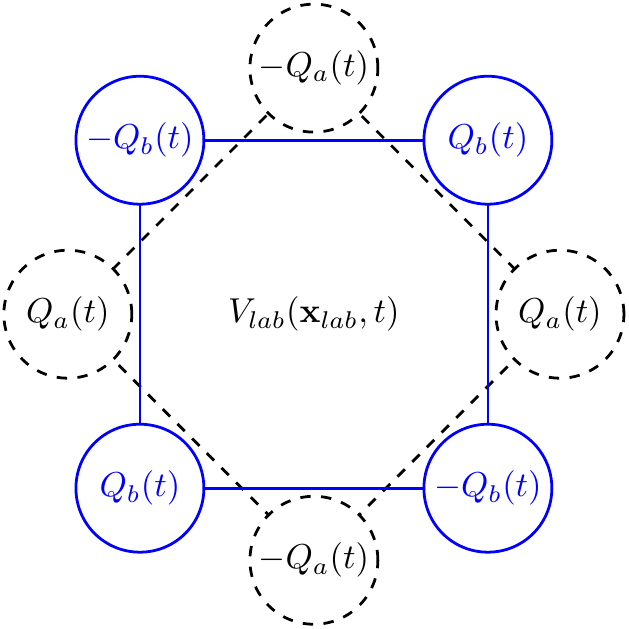}
\end{center}
\caption{Sketch of an experimental setup (top view) that can be used to generate the rotating saddle potential. Two sets of mechanically fixed electrodes ($a$ and $b$) are placed symmetrically around the center of the sample.  Time-dependent potentials $Q$ are applied to the electrodes with the indicated strengths and relative signs so that each set generate a quadrupole field in the center. When the amplitudes between the two sets have a $\pi/2$-phase shift, e.g. $Q_a(t) = Q \cos(2 \Omega t)$ and $Q_b(t) = Q \sin(2 \Omega t)$, the potential in the center $V(\mathbf{x}_{lab},t)$ becomes that of a rotating saddle.
\label{fig:setup}}
\end{figure}

{\it Possible application in graphene and on surface states of 3D topological insulators.} A rotating saddle potential can in principle easily be generated by a superposition of two oscillating quadrupole fields, as sketched in Fig.~\ref{fig:setup}. It is also interesting to note that a static disorder potential generates a rotationally invariant potential profile on average in the rotating frame. If the driving frequency is large enough this first order result is the dominating contribution. We therefore expect that the bound states are only weakly affected by disorder and imperfections.

Experimentally the localized states can be probed using scanning tunneling spectroscopy or via transport through the bound states. For clear experimental signatures of the trapped states the typical energy separation between neighboring $j$-values near the saddle center  $\hslash |\Omega|$ should at least exceed the temperature scale $T$. Experimentally frequencies above $\sim 10 \, \text{GHz}$ are very difficult to achieve with conventional radio frequency methods, limiting the temperature range of such setups to the sub-Kelvin scale. In the optical frequency range the temperature is no longer an issue. In this case the design of an appropriate laser setup to achieve the desired inhomogeneous field configuration is however more challenging. 

To summarize we propose to use a rotating saddle-shaped potential to trap massless 2D Dirac particles. This has potential applications in graphene, three-dimensional topological insulator surface states, and cold atom systems. Our study also demonstrates the richness of particle dynamics in external time- and space-dependent fields.

\acknowledgments

I'd like to thank C.-Y. Hou for useful comments on the manuscript and the Swedish research council (Vetenskapsr{\aa}det) for funding.

\bibliography{rotating_saddle}

\end{document}